\documentclass[iop]{emulateapj}

\def\lapp{\ifmmode\stackrel{<}{_{\sim}}\else$\stackrel{<}{_{\sim}}$\fi}
\def\gapp{\ifmmode\stackrel{>}{_{\sim}}\else$\stackrel{>}{_{\sim}}$\fi}

\usepackage{xspace}

\newcommand{\presto}{\texttt{PRESTO}\xspace}
\newcommand{\rfifind}{\texttt{rfifind}\xspace}
\newcommand{\prepsub}{\texttt{prepsubband}\xspace}
\newcommand{\accelsearch}{\texttt{accelsearch}\xspace}
\newcommand{\prepfold}{\texttt{prepfold}\xspace}
\newcommand{\spsearch}{\texttt{single\_pulse\_search.py}\xspace}
\newcommand{\nemodel}{\texttt{NE2001}\xspace}
\newcommand{\tempo}{\texttt{TEMPO}\xspace}

\begin{document}

\title{Constraining Radio Emission from Magnetars}

\author{
P.~Lazarus\altaffilmark{1},
V.~M.~Kaspi\altaffilmark{1,2},
D.~J.~Champion\altaffilmark{3},
J.~W.~T.~Hessels\altaffilmark{4,5},
R.~Dib\altaffilmark{1}
}

\altaffiltext{1}{Department of Physics, Rutherford Physics Building,
McGill University, 3600 University Street, Montreal, Quebec,
H3A 2T8, Canada; plazar@physics.mcgill.ca}

\altaffiltext{2}{Lorne Trottier Chair; Canada Research Chair}

\altaffiltext{3}{Max-Planck-Institut f\"ur Radioastronomie, Auf dem H\"ugel 69, 53121 Bonn, Germany}

\altaffiltext{4}{Netherlands Institute for Radio Astronomy
(ASTRON), Postbus 2, 7990 AA Dwingeloo, The Netherlands}

\altaffiltext{5}{Astronomical Institute ``Anton Pannekoek,''
University of Amsterdam, 1098 SJ Amsterdam, The Netherlands}

\begin{abstract}
We report on radio observations of five magnetars and two magnetar candidates carried out at 1950 MHz with the Green Bank Telescope in 2006-2007. The data from these observations were searched for periodic emission and bright single pulses. Also, monitoring observations of magnetar 4U~0142+61 following its 2006 X-ray bursts were obtained. No radio emission was detected was detected for any of our targets. The non-detections allow us to place luminosity upper limits of $L_{1950}$~\lapp~1.60 mJy kpc$^2$ for periodic emission and $L_{1950,~\mathrm{single}}$~\lapp~7.6 Jy kpc$^2$ for single pulse emission. These are the most stringent limits yet for the magnetars observed. The resulting luminosity upper limits together with previous results are discussed, as is the importance of further radio observations of radio-loud and radio-quiet magnetars.
\end{abstract}


\section{Introduction}
Magnetars are a subclass of pulsars powered by the decay of their ultra-strong magnetic fields, typically $B \sim 10^{14}$ to $10^{15}$ G (see \cite{wt06}, or \cite{mer08} for recent reviews). The X-ray emission of magnetars exhibits phenomena not seen in X-ray-detected rotation-powered pulsars. Magnetars typically have a greater X-ray luminosity than can be explained by their spin-down luminosity (i.e. $L_X > \dot{E}$), unlike rotation-powered pulsars. In addition, magnetars are very variable objects; they have a wide variety of X-ray emission behavior such as short single ($\sim100$ ms) bursts, periods of outburst containing many short bursts, giant flares lasting hundreds of seconds and which exhibit pulsations in their fading tails, flux enhancements lasting hundreds of days and X-ray pulse profile variations \citep[see][for more details]{wt06, kas07, re11}. Magnetars also have complicated timing properties such as timing noise, torque variations, and glitches \citep[e.g.][]{klc00, dis+03, dkg07, dkg08, dib09}.

According to the McGill SGR/AXP Online Catalog\footnote{http://www.physics.mcgill.ca/$\sim$pulsar/magnetar/main.html, as of 2011 August 25}, there are currently 16 confirmed magnetars, consisting of seven Soft Gamma Repeaters (SGRs) and nine Anomalous X-ray Pulsars (AXPs), as well as seven magnetar candidates (four SGR and three AXP candidates). These numbers are small relative to the total number of known rotation-powered pulsars, nearly 2000, according to the online ATNF Pulsar Catalog\footnote{http://www.atnf.csiro.au/research/pulsar/psrcat/} \citep{mhth05}.

Prior to 2006, there was no firm detection of pulsed radio emission from a magnetar. Since that time, three magnetars have been found to have emission at radio frequencies: XTE~J1810$-$197, 1E~1547.0$-$5408, and PSR~J1622$-$4950 \citep{crh+06, crhr07, lbb+10}. The radio emission of these three sources has properties that are common among them, but which are not shared with the bulk of the rotation-powered radio pulsar population. Specifically, the three radio magnetars show variable spectral indices, variable pulse profiles, and variable radio luminosities. The spectra observed from radio-detected magnetars (even when taking into consideration the variability observed in their spectral indices) are generally very flat, or rising with observing frequency, in contrast to the majority of rotation-powered pulsars, whose spectra steeply decline with observing frequency, with mean spectral index $\langle \alpha \rangle = -1.8 \pm 0.2$ \citep{mkkw00}. The origin of the spectral flatness in radio emission from magnetars is not known.

The discoveries of radio-loud high-magnetic-field pulsars and magnetars have spurred observations of other magnetars in the hope of detecting radio emission \citep{bri+06, chk07}. Limits of $S_{1400} \lapp 20~\mu$Jy have been placed on three Southern magnetars, 1RXS~J170849.0$-$400910, 1E~1841$-$045 and 1E~1048.1$-$5937, and one magnetar candidate, AX~J1845$-$0258 \citep{bri+06, chk07}. Also, upper limits on the flux density of single pulses at 1400 MHz from the above-listed four sources have been placed in the range 0.9 -- 1.1 Jy \citep{chk07}.

In this paper we describe similar radio observations of magnetars in the Northern sky. The goal was to discover more magnetars that exhibit pulsed radio emission. Increasing the number of radio-loud magnetars known could offer insight into the pulsar emission mechanism, the behavior of matter in ultra-strong magnetic fields, as well as the possible evolutionary relationship between magnetars and the much larger known population of rotation-powered pulsars. 

This paper is organized as follows: \S \ref{sec:observations} describes the observations undertaken. The analysis performed on the data collected is described in \S \ref{sec:analysis}. The results are presented in \S \ref{sec:results}. In \S \ref{sec:discussion} we put the results into context and discuss their implications.
 
\section{Observations}
\label{sec:observations}
Observations of five confirmed magnetars and two magnetar candidates were carried out using the NRAO 100-m Green Bank Telescope\footnote{http://www.gb.nrao.edu/gbt/} in 2006-2007. The goal was to observe the magnetars and magnetar candidates visible from the GBT in order to  detect radio emission, or in the absence of a detection, establish baseline measurements should a target turn on as a radio source some time in the future. The proposal also included Target-of-Opportunity (ToO) observations to be triggered if a source exhibited an outburst, as determined by on-going X-ray monitoring observations.

Total intensity data were recorded by the GBT's Pulsar Spigot backend, an auto-correlation spectrometer \citep{kel+05}. Lagged-products were converted to spectra off-line. The result is spectra containing 600 MHz of usable bandwidth centered at 1950 MHz (S-band) divided evenly into 768 channels, written out every 81.92 $\mu$s.

\subsection{Targets Observed}
Here we present relevant details of our seven targets. In total, 19 observations of a variety of durations were made between 2006 November and 2007 October as summarized in Table \ref{tab:observations}. For each source we present its best distance estimate, together with an estimate of the free electrons along the line-of-sight toward the source from the \cite{cl02} model. The free electron content is parameterized by the dispersion measure (DM). Estimates of the model are used to determine the upper limit on DM searched for each target. In all cases the DM searched was \gapp 2 times larger then the maximum DM predicted along the line-of-sight.

\textbf{AXP 1E~1841$-$045}: This source has a rotational period of 11.8 s \citep{vg97}. The estimated distance to this source is $8.5^{+1.3}_{-1}$ kpc \citep{tl08}, which, together with its position, suggests a DM $\simeq$ 800 cm$^{-3}$ pc \citep{cl02}. The supernova remnant\footnote{Kes 73 is also referred to as G27.4+0.0 in Green's online Catalogue of Galactic Supernova Remnants: http://www.mrao.cam.ac.uk/surveys/snrs/ \citep{gre09}.} Kes 73 is associated with 1E~1841$-$045. The pulsar is monitored regularly by NASA's \emph{Rossi X-ray Timing Explorer}\footnote{Information about \emph{RXTE} monitoring targets and observations can be found on the telescope's website: http://heasarc.gsfc.nasa.gov/docs/xte/.} \citep[\emph{RXTE}, ][]{dkg08, dib09}. The analysis of these data allow for an accurate rotational ephemeris for the pulsar to be determined for our GBT observing epoch; this is shown in Table \ref{tab:ephems}.

\textbf{1E~2259+586}: The magnetar has a period of 6.98 s \citep{fg81}. The source is estimated to be at a distance of $4.0\pm0.8$ kpc \citep{tll10}. This distance and the source position correspond to a line-of-sight DM $\simeq$ 150 cm$^{-3}$ pc. 1E~2259+586 is associated with the supernova remnant\footnote{CTB~109 is also known as G109.1-1.0 in Green's catalogue.} CTB~109. 1E~2259+586 is also monitored regularly using \emph{RXTE} \citep{dib09}, and the resulting ephemeris is shown in Table \ref{tab:ephems}.

\textbf{4U~0142+61}: This source has a period of 8.69 s. The distance to this source is estimated to be $3.6\pm0.4$ kpc \citep{dk06}. This distance suggests a DM $\simeq$ 100 cm$^{-3}$ pc. The rotational ephemeris for 4U~0142+61 for our observing epoch is shown in Table \ref{tab:ephems}. This ephemeris was also computed using \emph{RXTE} data from regular monitoring observations \citep{dkg07}.

Within a few months before and after the start of our GBT observations, X-ray bursts were detected from 4U~0142+61 with \emph{RXTE} \citep{gdkw07,gdk11}. The six bursts were detected in three separate observations on MJDs 53831 (one burst), 53911 (four bursts) and 54138 (one burst). The last of the six bursts had the largest peak X-ray flux. Monitoring observations of 4U~0142+61 were made on the day of this largest burst, as well as at 1 day, 1 week, 3 weeks, 2 months and 8 months following this burst (see Table \ref{tab:observations}).

\textbf{SGR~1806$-$20}: The source has a rotational period of 7.56 s \citep{kds+98} and is estimated to be at distance of $8.7^{+1.8}_{-1.5}$ kpc \citep{bcfc08}, which suggests DM $\simeq$ 750 cm$^{-3}$ pc. SGR~1806$-$20 was observed to have a giant flare in late 2004 \citep{hbs+05}. Following the flare, a fading, extended, unpulsed radio nebula was observed around the source \citep{gkg+05}. Two days after the flare, a search for pulsed radio emission was performed at 1400 MHz using the Parkes Radio telescope. No pulsed emission was detected down to a limit of $S_{1400} \sim 0.2$ mJy \citep{gkg+05}.

\textbf{SGR~1900+14}: This SGR has a periodicity of 5.17 s \citep{ksh99}. Its distance is estimated to be 12 -- 15 kpc, which yields an estimated DM $\simeq$ 700 cm$^{-3}$ pc, using the upper limit of the distance estimate. SGR~1900+14 was observed to have a giant flare in mid-1999 \citep{hcm+99}. Unpulsed, fading radio emission was also observed around SGR~1900+14 following the flare \citep{fkb99}.

\textbf{AX~J1845$-$0258}: This magnetar candidate was detected in a single 1993 \emph{Advanced Satellite for Cosmology and Astrophysics} (\emph{ASCA}) observation with $P = 6.97$~s \citep{tkk+98,gv98}. AX~J1845$-$0258 is tentatively classified as an AXP candidate because of its long period, soft spectrum, and low luminosity. The distance to the pulsar is estimated to be 8.5~kpc \citep{tkk+98}. This distance corresponds to a DM $\simeq$ 750 cm$^{-3}$ pc. AX~J1845$-$0258 is spatially coincident with the supernova remnant G29.6+0.1 \citep{ggv99}, suggesting that the remnant and magnetar candidate are associated. No subsequent detection of this pulsar has been made \citep[see][for example]{tkgg06}, hence we have only an approximate period estimate.

\textbf{GRB~050925}: This $\gamma$-ray burst (GRB) was detected as a short burst (104 ms) with the \emph{Swift} satellite's burst alert telescope (BAT) on Sept. 25, 2005 \citep{hbb+05,sbb+11}. GRB~050925 is considered an SGR candidate because of its soft X-ray spectrum, which is uncharacteristic of short GRBs, and its localization in the Galactic plane ($l = 72.32(3)^\circ$, $b = -0.09(3)^\circ$). Within the BAT error circle two variable X-ray sources are plausible counterparts of GRB~050925 \citep{sbb+11}. GRB~050925 was also followed-up at 4900 MHz with the Westerbork Synthesis Radio Telescope. No sources were detected within the BAT error circle with $S_{4900} > 76 \mu$Jy, and one constant radio source was detected just outside the error circle \citep{hor05a, hor05b}. This magnetar candidate is still unconfirmed, without a period measurement, or an estimate of its magnetic field strength. Nevertheless we observed the position of GRB~050925 to search for radio bursts or pulses from this possible magnetar. For the purposes of this work we use the distance estimate from \cite{sbb+11}, 8.8~kpc, which assumes GRB~050925 is associated with the H$_{\mathrm{II}}$ region W~58. This distance and line-of-sight correspond to a DM estimate of 300 cm$^{-3}$~pc \citep{cl02}.

The properties described above are summarized in Table \ref{tab:sourcedetails}.

\section{Data Analysis}
\label{sec:analysis}
Data were searched for both periodic signals and for bright individual pulses using the \presto suite of pulsar search programs\footnote{http://www.cv.nrao.edu/$\sim$sransom/presto/} \citep{ran01}. Masks to remove narrow-band impulsive and periodic radio frequency interference (RFI) were created using \presto's \rfifind program. The masks produced by \rfifind were adjusted slightly by hand to mask any additional bad frequency channels or time intervals from the data. These masks were applied to the data before dedispersion. 

In seven data sets, the mean level had a large step during the observation. This was a result of large bursts of broadband RFI, which caused offsets in the mean level of GBT Pulsar Spigot data (S. Ransom, private communication). This behavior has also been seen in other Spigot data sets \citep[e.g.][at 820 MHz]{kml+09}. Our observations that were affected are: 4U~0142+61 (MJDs 54056, 54059, 54146), 1E~1841$-$045 (MJD 54052), 1E~2259+586 (MJD 54059), and SGR~1806$-$20 (MJDs 54190, 54211). To deal with these jumps we did the following. If the jump in the data mean occurred near the beginning or end of an observation, only the longer portion of the observation was analysed. The integration time of the observation was adjusted accordingly. Alternatively, if the jump occurred near the middle of an observation, the observation was split into two portions and each was analysed independently. 

Data were dedispersed according to the following plan. For DMs in the range 0 to 1247 cm$^{-3}$~pc, DM step sizes of $\Delta$DM = 1 cm$^{-3}$~pc and downsampling by a factor of 4 were used. In the range DM = 1248 to 2446 cm$^{-3}$~pc, data were downsampled by a factor of 8 and the $\Delta$DM used was 2 cm$^{-3}$~pc. Finally, for DMs larger than 2448 cm$^{-3}$~pc, $\Delta$DM = 3 cm$^{-3}$~pc was used and data were downsampled by a factor of 16. Maximum DM values used varied by target, depending on the prediction for the given line-of-sight by the \nemodel model \citep[][see \S \ref{sec:observations}]{cl02}. Table \ref{tab:sourcedetails} shows the maximum DM values searched for each target. Dedispersed time series were produced using \presto's \prepsub. 

The resulting time series were inspected for steady periodic signals using \presto's \accelsearch, which searches for peaks in Fourier transforms. Sensitivity to narrow pulses was increased by summing up to 16 harmonics. The search for periodic signals was repeated with a red-noise removal technique, which included subtraction of a running median from the time series.

For the three magnetars with known rotational ephemerides (see Table \ref{tab:ephems}), the data were folded at the known ephemeris to search for periodic emission. Even when an ephemeris was available, the dedispersed time series were still searched using \accelsearch for other periodic sources in the field-of-view.

Significant periodicity candidates were folded using the DM trial for which the candidate's period had the largest signal-to-noise ratio. Folding was performed using \prepfold. The resulting plots were examined by eye. 

Bright single pulses were searched for using \presto's \spsearch, which uses matched filtering with top-hat filters having widths ranging from 0.33 to 50 ms. The data were then downsampled by a factor of 4 and re-searched to increase sensitivity to single pulses with durations between 1.3 and 200 ms. The range of pulse durations the analysis is sensitive to is reasonable given the pulses observed from rotating radio transients \citep[see, e.g.][]{mll+06}. For each observation, single pulse events with signal-to-noise ratio $\mathrm{(S/N)} \geq 8$ were plotted and examined by eye.

\section{Results}
\label{sec:results}
None of the observations listed in Table \ref{tab:observations} resulted in the detection of periodic or impulsive radio emission. The observations were used to place upper limits on any such emission, as we describe next.

\subsection{Upper Limits on Periodic Emission}
In order to compute upper limits on periodic emission for the observations analyzed in this work, we used the modified radiometer equation \citep[e.g.][]{lorimer+kramer04},

\begin{equation}
    \label{eq:radiometer}
    S_{min} = \beta  \mathrm{(S/N)}_{min}
                \frac{\left [ \left ( T_{rcv} + T_{sky} \right ) / G + S_{SNR} \right ]}
                    {\sqrt{n_p t_{int} \Delta f}} 
                \sqrt{\frac{\delta}{1-\delta}},
\end{equation}

\noindent where $S_{min}$ is the minimum detectable flux density in mJy, $\beta$ is the signal degradation factor due to quantization, $\mathrm{(S/N)}_{min}$ is the minimum signal-to-noise ratio considered, $T_{rcv}$ and $T_{sky}$ are the receiver and sky temperatures in K, respectively, $G$ is the telescope gain in K Jy$^{-1}$, $S_{SNR}$ is the flux density of the supernova remnant in Jy (if there is such an association), $n_p$ is the number of polarisations summed, $t_{int}$ is the integration time in seconds, $\Delta f$ is the observing bandwidth in MHz, and $\delta$ is the assumed duty cycle, ranging between 0 and 1. Both the integration time and observing bandwidth were adjusted downward to take into consideration data masked due to RFI removal.

The duty cycle is related to the width according to $\delta = W_b/P$, however this width is not the intrinsic width of the pulsar's integrated profile. The intrinsic width is effectively broadened by the finite sampling time, $t_{samp}$, dispersive smearing within each channel, $t_{DM}$, and multi-path scattering with the ISM, $t_{scatt}$. The broadened width is related to the intrinsic width, $W_i$, according to

\begin{equation}
    \label{eq:width}
    W_b = \sqrt{W_i^2 + t_{samp}^2 + t_{DM}^2 + t_{scatt}^2}.
\end{equation}

\noindent Here, the scattering time, $t_{scatt}$, depends on the degree of inhomogeneity of the free electrons along the line-of-sight, which is also predicted using the \nemodel model \citep[details can be found in][]{cl02}. 

For the purposes of this work, the signal degradation factor due to quantization, $\beta$, is $\simeq 1$ since 16 bits are used. The gain of the GBT, at 1950 MHz, is\footnote{See the GBT proposer's guide, http://www.gb.nrao.edu/gbtprops/man/GBTpg.pdf.} $G = 1.9$ K Jy$^{-1}$. Also, the GBT's S-band receiver system is maintained at a temperature of $T_{rcv} = 20$ K. The sky temperature, $T_{sky}$, is the sum of contributions due to Galactic synchrotron radiation, as determined from the \cite{hssw82} all-sky radio map\footnote{The electronic HEALPix version of the \cite{hssw82} 408~MHz map provided by NASA LAMBDA was used: http://lambda.gsfc.nasa.gov/product/forground/fg\_haslam\_get.cfm. The Galactic synchrotron emission is assumed to have a power-law spectrum with index $-2.5$ \citep{kfl+11}.}, plus the 2.73 K contribution from the cosmic microwave background. The supernova remnant associations and their respective flux densities are shown in Table \ref{tab:sourcedetails}. Finally, intrinsic pulse widths are taken to be between 3\% and 50\% of the pulse period. These widths are then broadened according to Equation \ref{eq:width}.

To determine the minimum detectable signal-to-noise ratio, $\mathrm{(S/N)}_{min}$, in the presence of RFI, an observation of a known pulsar, PSR~J1907+0918, was used. Subsets of the observation of various durations were folded. The resulting folded profiles were examined by eye to determine an approximate threshold of detectibility, $\mathrm{(S/N)}_{min} = 4$. Assuming purely white noise, a 4 $\sigma$ detection has a probability of $P\left( 4~\sigma \right) \simeq 6.3 \times 10^{-5}$ of occuring by chance. 

For magnetars for which a current ephemeris was available (1E~1841$-$045, 1E~2259+586 and 4U~0142+61; see Table \ref{tab:ephems}), $\mathrm{(S/N)}_{min} = 4$ was used since the exact period at the time of the observation was known. 

For sources for which there was an uncertainty on the folding frequency (SGR~1806$-$20, SGR~1900+14, and AX~J1845$-$0258), the value of $\mathrm{(S/N)}_{min}$ used was larger, because of the larger number of Fourier bins searched. The number of bins searched was computed by estimating the frequency at the epoch of the observation, which was extrapolated using a previously published frequency and frequency derivative, or in the case of AX~J1845$-$0258, a previously published frequency and an assumed magnetic field strength equal to $5 \times 10^{15}$~G (i.e. twice the largest inferred magnetic field strength for any known magnetar). A conservative fractional uncertainty on the frequency of $10^{-4}$ was used for all three sources. This was done to take into consideration the possibility of a very large glitch, or anti-glitch. In all cases, the change in frequency due to a putative glitch was comparable to, or dominated by, the uncertainty in the extrapolated frequency. The frequency range searched was conservatively taken to be 3 times the change in frequency due to a potential glitch or anti-glitch, as described above. This range was divided by the frequency resolution, $1/t_{int}$, to find the number of Fourier bins searched. This is equivalent to the number of independent trials. In the case of SGR~1806$-$20 and SGR~1900+14, the frequency range searched still amounted to only one Fourier bin. In the case of AX~J1845$-$0258, 100 Fourier bins were searched.

For magnetar candidate GRB~050925, there is no information about the spin period of the potential neutron star. Therefore, all Fourier bins were searched (i.e. 5493164 and 4211426 bins for the observations on MJDs 54053 and 54056, respectively).

The probability corresponding to $4~\sigma$ was divided by the number of Fourier bins searched and converted back to an equivalent Gaussian sigma, which was used in Equation \ref{eq:radiometer}. The values of $\mathrm{(S/N)}_{min}$ used in Equation \ref{eq:radiometer} are 5 for the observation of AX~J1845$-$0258, and 6.79 and 6.75 for the observations of GRB~050925 on MJDs 54053 and 54056, respectively. 

\begin{figure}[h]
    \centering
    \includegraphics[scale=0.7]{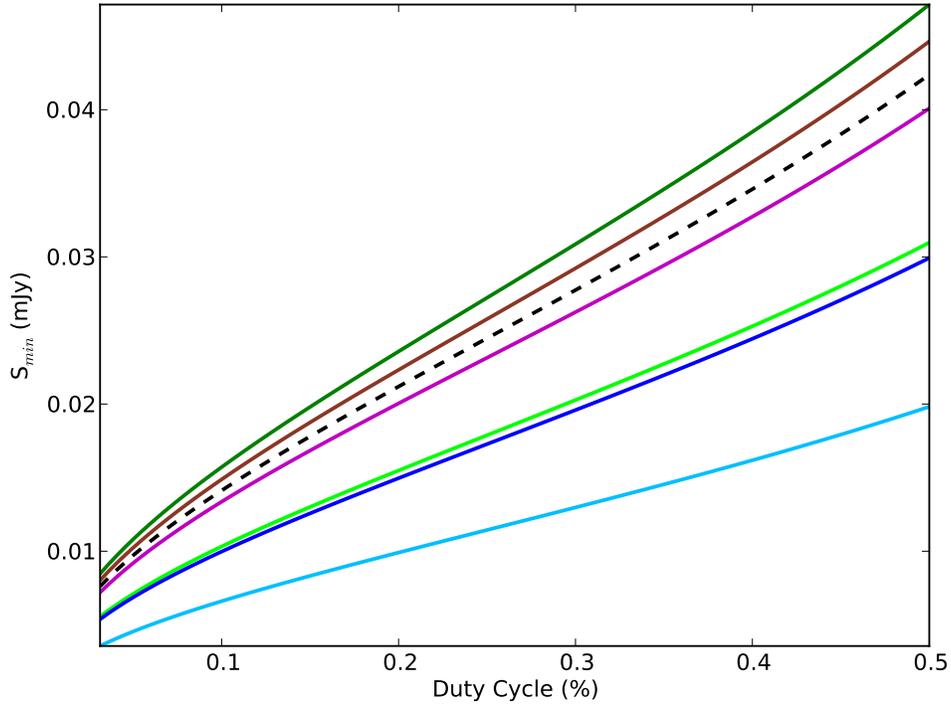}
    \caption{Best flux density limits at 1950 MHz as a function of duty cycle for periodic emission from each of the seven targets observed. From top to bottom the curves correspond to 1E~2259+586, 1E~1841$-$045, GRB~050925 (dashed), AX~J1845$-$0258, SGR~1900+14, SGR~1806$-$20, and 4U~0142+61. \label{fig:periodicity uplim}}
\end{figure}

For each observation, an upper limit on the flux density of periodic emission has been computed as a function of pulse width. See Figure \ref{fig:periodicity uplim} for an example of the dependence of the upper limit on pulse width. Assuming a duty cycle of $\delta = 5\%$, all observations have a minimum detectable flux density of $S_{min} \lapp$ 0.013 mJy or lower. By assuming a distance for each source, this can be translated to a minimum detectable luminosity of L$_{min} \lapp$ 1.6 mJy~kpc$^2$ for all observations analysed. Table \ref{tab:uplim} summarizes the results\footnote{Note that the number of entries in Table \ref{tab:uplim} exceeds the number of entries listed in Table \ref{tab:observations} because two observations were split due to RFI-induced jumps in the baseline.}.

\subsection{Upper Limits on Single Pulses}
In addition to searching for periodic signals, the data were also searched for dispersed bright single pulses. Many single pulse events with signal-to-noise ratio $>$ 8 were detected, however none was found to be consistent with an astrophysical origin. 

\begin{figure}[h]
    \centering
    \includegraphics[scale=0.7]{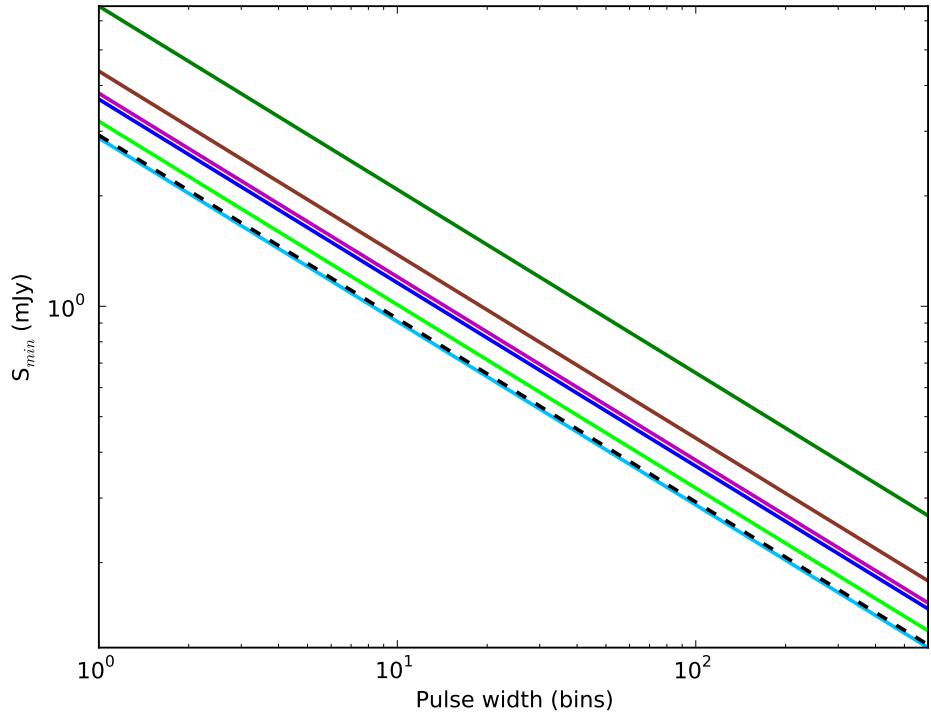}
    \caption{The best flux density limits for single pulses at 1950 MHz as a function of pulse width for each of the seven targets observed. From top to bottom the curves correspond to 1E~2259+586, 1E~1841$-$045, AX~J1845$-$0258, SGR~1806$-$20, SGR~1900+14, GRB~050925 (dashed), and 4U~0142+61. \label{fig:splim}}
\end{figure}

To compute the minimum detectable flux density of single pulses in the observations, the formalism introduced in \cite{cm03} was used. The relation between the brightness of a single pulse of astrophysical origin and its measured signal-to-noise ratio is given by

\begin{equation}
    \label{eq:splim}
    S_i = \frac{ \left ( \mathrm{S/N} \right )_b S_{sys}}{W_i} \sqrt{\frac{W_b}{n_p \Delta f}},
\end{equation}

\noindent where $S_i$ is the intrinsic flux density of a pulse of width $W_i$, (S/N)$_b$ is the broadened signal-to-noise ratio, as measured by matched filtering, $S_{sys} = (T_{rcv} + T_{sky})/G + S_{SNR}$ is the system equivalent flux density, $W_b$ is the broadened width of the single pulse. $W_i$ and $W_b$ are related by Equation \ref{eq:width}. In this work, the minimum signal-to-noise ratio considered for a single pulse is conservatively chosen to be $\mathrm{(S/N)}_{b, min} = 8$. The large value of $\mathrm{(S/N)}_{b, min}$ used is meant to exclude not only noise, but some RFI as well. For each observation, the minimum detectable luminosity of a 10-ms single pulse is reported in Table \ref{tab:uplim}. Examples of the dependence of $S_{min}$ on pulse duration are shown in Figure \ref{fig:splim}.

Based on the previous discussion and the fact that no astrophysical single pulses were detected, we could conclude that the sources observed do not emit single pulses brighter than $L_{min,~single}$, the minimum detectable single pulse luminosity. However, it is also possible the sources in question \emph{do} occasionally emit single pulses bright enough to be detected, but the rates at which these pulses are emitted are sufficiently small that no pulses were detected in our observations. With this in mind, a 3$\sigma$ upper limit can be placed on the single pulse rate, knowing that $< 1$ single pulses were detected and assuming such pulses are emitted randomly. We have used the sum of all unmasked integration time for each source to compute limits on the single pulse rate. These apply only to pulses bright enough to be detected with durations between 0.33 ms and 200 ms, and are shown in Table \ref{tab:sprate}. 

\subsection{ToO Observations of 4U~0142+61}
\begin{figure}[h]
    \centering
    \includegraphics[scale=0.7]{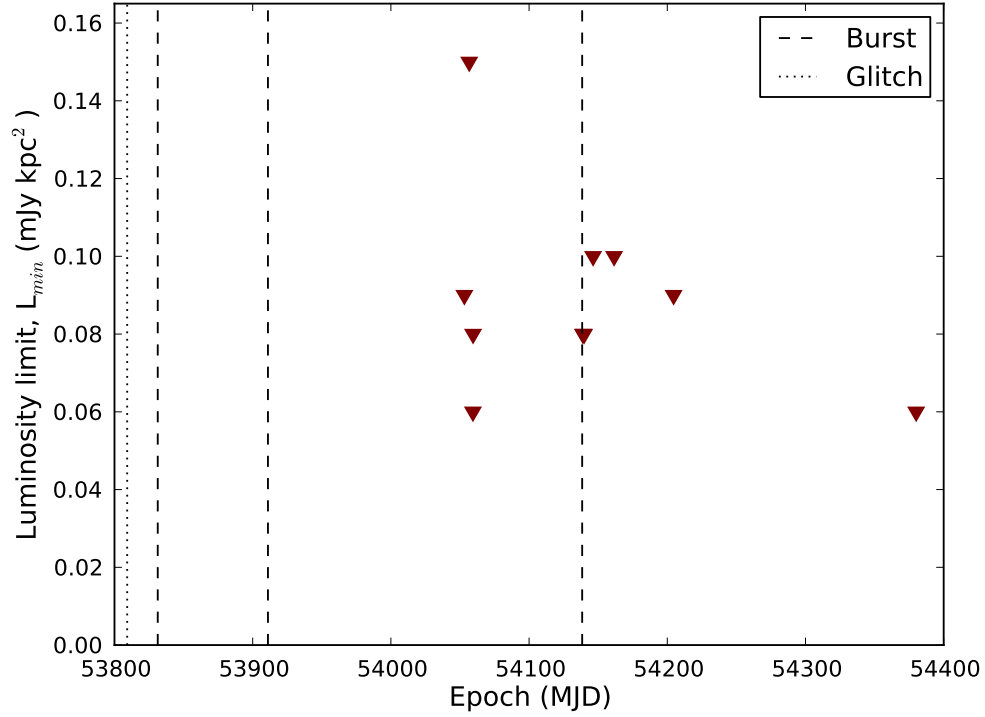}
    \caption{A timeline of activity for 4U~0142+61 including a glitch (dotted line), six X-ray bursts at three different epochs \citep[dashed lines, from][]{gdkw07,gdk11}, and upper limits on radio emission at 1950 MHz (downward-pointing triangles). Note that typical luminosities for the radio-detected magnetars are significantly off the top of the plot \citep{crh+06, crhr07, lbb+10}. \label{fig:0142timeline}}
\end{figure}

ToO observations were triggered when an X-ray burst from 4U~0142+61 was detected on MJD 54138 using \emph{RXTE} \citep{gdkw07,gdk11}. The first of six GBT ToO observations were on the day the burst was detected. Additional observations were scheduled at increasing intervals following the burst. The spacing between subsequent observations were: 1 day, 1 week, 2 weeks, 1 month, and 6 months. None of these observations resulted in a detection of emission from 4U~0142+61. The collection of radio observations of 4U~0142+61, and a timeline of its activity in 2006 are shown in Figure \ref{fig:0142timeline}.

\subsection{Constraints on Other Pulsars in the Field-of-View}
The surface density of known pulsars on the sky is large enough that a chance inclusion of an unrelated source in an observation is possible. We therefore also searched for signals at periods far from those of the magnetar targets.

No new pulsars were discovered in any of the observations searched. However, one previously known pulsar, PSR~J1907+0918, a 226-ms pulsar with a DM = 357 cm$^{-3}$ pc \citep{lx00}, was detected in our only observation of SGR~1900+14, on MJD 54053.

\begin{figure}[h]
    \centering
    \includegraphics[scale=0.7]{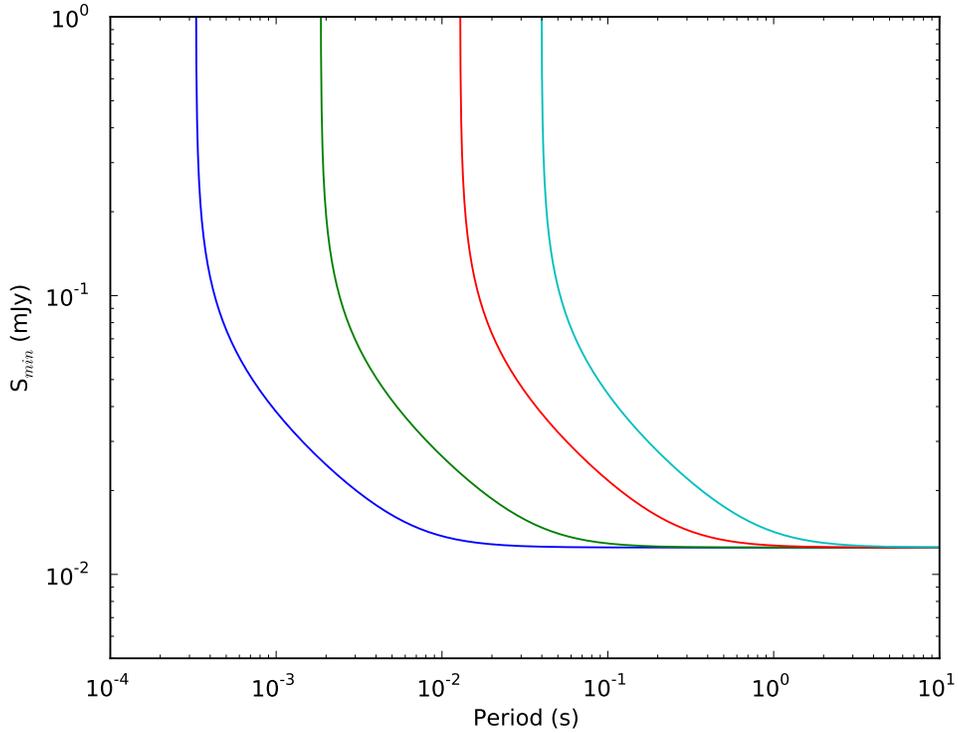}
    \caption{Typical flux density limits for a blind search for pulsars as a function of spin period for various DMs. These limits are for the observation of AX~J1845$-$0258 on MJD 54053. An intrinsic duty cycle of 5\% is assumed. The duty cycle is broadened by the observing system's finite sample time, dispersive smearing and scattering. From left to right the lines correspond to dispersion measures of 0 cm$^{-3}$ pc, 500 cm$^{-3}$ pc, 1000 cm$^{-3}$ pc, and 2000 cm$^{-3}$ pc.\label{fig:blindlim}}
\end{figure}

The minimum detectable flux density for an unknown pulsar as a function of spin period can be computed for the blind searches performed. Representative values are computed using Equation \ref{eq:radiometer}, assuming a duty cycle of $\delta =$ 5\%, a period of $P = 100$~ms and DM = 100 cm$^{-3}$~pc. The limits of the blind searches vary among targets since $T_{sky}$ and $S_{SNR}$ affect the system temperature. Also, the integration time and the RFI conditions (i.e. amount of data masked) are different for each observation. Typical results are reported in Table \ref{tab:uplim} and the dependences on DM and pulse period are shown in Figure \ref{fig:blindlim}.

\section{Discussion}
\label{sec:discussion}
This work was designed to detect radio emission from a collection of magnetars and magnetar candidates at 1950 MHz. The results presented here are more constraining than previously published limits at 1400 MHz for the same sources \citep{bri+06,chk07,gkg+05,dkh+07}\footnote{Because of the large difference in observing frequencies, we do not directly compare our results with the limits obtained by \cite{lx00} for SGR~1900+14 at 430 MHz.}. However, considering the variable nature of the three known radio-loud magnetars, the results of this work are also complimentary to the previously published constraints. Even though the variable nature of the radio-loud magnetars limits the conclusions that can be drawn from individual non-detections, it is still worthwhile to consider the ensemble of non-detections.

The three known radio-loud magnetars have properties different from those of most other radio pulsars. Most importantly, the known radio magnetars are transient, i.e. their emission is often absent. Examples of this behavior include the sudden appearance then later fading of XTE~J1810$-$197's radio emission over the course of $\sim$ 250 days \citep{ccr+07}, as well as the more sporadic behavior observed in 1E~1547.0$-$5408, which, for example, was not detected in observations on 2009 January 22 and 2009 January 23, but was detected two days later on 2009 January 25 \citep{chr09, bip+09}. Also, when the known radio magnetars are observed to be ``on'', their luminosities are variable by a factor of a few \citep[e.g.][]{ccr+07, lbb+10}. Therefore, the results presented in this work cannot be used to provide constraints when the sources were not being observed.

\subsection{Beaming Fractions of Magnetars}
Slowly rotating pulsars have relatively large light cylinder radii and small polar caps, and thus are expected to have narrow radio beams \citep[e.g.][and references therein]{tm98}. Therefore, a magnetar may not be detected at radio frequencies because its radio beam does not intersect the Earth. Unfortunately, it is difficult to estimate the fraction of the celestial sphere illuminated by any given pulsar without knowing the geometry involved. The X-ray pulse profiles of magnetars are typically broad \citep{wt06}, thus their X-ray beaming fractions are likely quite large. Therefore, it is unreasonable to make assumptions concerning the geometries of known magnetars based on their X-ray detections. Instead, naively assuming that the radio beaming properties of magnetars are similar to those of rotation-powered radio pulsars, we use empirical results for the beaming fraction of radio pulsars to attempt to constrain that of magnetars \citep{tm98}. 

\cite{tm98} estimate the fraction of pulsars, $f$, beamed towards the Earth as a function of spin period, $P$, to be

\begin{equation}
    \label{eq:beamingfrac}
    f \left( P \right) = 0.09 \left[ \log \left( \frac{P}{\mathrm{s}} \right) - 1 \right]^2 + 0.03.
\end{equation}

Given that there are three known radio-loud magnetars out of a total of 13 sources observed \citep[this work;][]{bri+06, chk07, lek+09} it is possible to estimate the probability that at least one of the undetected sources is beamed towards the Earth. Using a representative period of 7~s, Equation \ref{eq:beamingfrac} can be used to compute the probability that 4 or more magnetars are beamed towards the Earth (that is, at least one of the undetected magnetars is beamed towards us). We find $P(\geq4) = 0.06\%$. This is equivalent to saying that if we had detected one of our targets, Equation \ref{eq:beamingfrac} would be inconsistent with the magnetar data at the 3.2 $\sigma$ level. Thus the non-detections reported here are consistent with being due to unfortunate beaming, assuming Gaussian statistics and that Equation \ref{eq:beamingfrac} applies.

However, if we assume Equation \ref{eq:beamingfrac} applies, and again consider representative period of 7 s, we find the probability of three or more magnetars being beamed towards us is $P(\geq3) \simeq 0.75\%$ ($\simeq 2.4 \sigma$), which is suggestive that radio-magnetars have wider beams than rotation-powered pulsars with similar periods. In fact, by recasting Equation \ref{eq:beamingfrac} in terms of the pulsar beam's half-opening angle, $\rho$, we find that $\rho \sim 30^{\circ}$ is required for three out of 13 sources to be beamed towards us to be consistent with random orientations at the 3$\sigma$ level. This value of $\rho$ is $\sim 10$ times larger than what is used by \cite{tm98}, $\rho \sim 2^{\circ}$, for $P$ = 7~s, and is much more accommodating of the large duty cycles of the radio profiles of 1E~1547.0$-$5408 and PSR~J1622$-$4950 \citep[14\% and $\sim$11\%, respectively;][]{crhr07, lbb+10}. This suggested difference in $\rho$ for magnetically powered and rotationally powered pulsars further reinforces the likely different origin of the radio emission in these objects.

\section{Conclusions}
\label{sec:conclusion}
Using the GBT's Pulsar Spigot, five magnetars and two magnetar candidates were observed at 1950 MHz over the course of $\sim$ 1 year. The data were searched for pulsed periodic emission, as well as bright single pulses. None of the targets observed were detected. The non-detections were used to place stringent constraints on the presence of any radio emission at the observing epochs. None of the seven targets observed showed periodic emission with luminosity larger than 1.6 mJy kpc$^2$, or single pulse emission with luminosity larger than 7.6 Jy kpc$^2$ at 1950 MHz in any of the data sets.

It is difficult to use the upper limits derived from this work to constrain the physical properties of the sources, since previous results suggest radio emission from magnetars is highly variable. It is possible that observations were merely scheduled, by chance, at times when the magnetars' radio emission was too faint to be detected. On the other hand, it is also possible that the non-detections are nothing more than unfortunate beaming geometries. However, this work and previous work have combined to detect three out of 13 magnetars observed at radio frequencies. This suggests that the radio beaming fraction of magnetars is larger than that of rotation-powered pulsars with similar periods, in agreement with lines of reasoning.

Finally, if any of the sources studied here becomes visible at radio frequencies in the future, our results will be invaluable for providing a baseline with which detections can be compared. This may help us understand what magnetospheric conditions are required to produce radio emission, and the details of plasmas and currents in magnetar magnetospheres. What is learned from magnetars could also likely help elucidate the radio emission mechanism for pulsars in general. For this reason, the known magnetars should continue to be observed at radio frequencies.

We thank the anonymous referee for useful comments. PL acknowleges support from an NSERC CGS-D award, and a Trottier Accelerator Award. VMK acknowledges support from an NSERC Discovery Grant, FQRNT via the Centre de Recherche en Astrophysique du Qu\'ebec, CIFAR, a Canada Research Chair and the Lorne Trottier Chair in Astrophysics and Cosmology, and by a Killam Research Fellowship. DJC acknowledges support from an NSERC postdoctoral fellowship and CSA supplement. JWTH is a Veni Fellow of the Netherlands Foundation for Scientific Research (NWO). The Robert C. Byrd Green Bank Telescope (GBT) is operated by the National Radio Astronomy Observatory which is a facility of the U.S. National Science Foundation operated under cooperative agreement by Associated Universities, Inc.

\clearpage

\begin{center}
\begin{deluxetable}{lccccc}
\tablecaption{Summary of GBT Observations of Five Magnetars and Two Magnetar Candidates. \label{tab:observations}}
\tablewidth{0pt}
\startdata        
        \hline
        \hline
        Source & Date & Epoch & RA & Decl. & Obs. length \\
        & & (MJD) & (J2000) & (J2000) & (s) \\
        \hline
        1E~1841$-$045 & November 13, 2006 & 54052.96 & 18:41:19 & $-$04:56:11 & 3600 \\
        \hline
        1E~2259+586 & November 14, 2006 & 54053.14 & 23:01:08 & 58:52:47 & 3600 \\
        & November 20, 2006 & 54059.23 & & & 10800 \\
        \hline
        4U~0142+61 & November 14, 2006 & 54053.18 & 01:46:22 & 61:45:06 & 2340 \\
        & November 17, 2006 & 54056.71 & & & 1500 \\
        & November 20, 2006 & 54059.35 & & & 16140 \\
        & February 7, 2007 & 54138.95\tablenotemark{a} & & & 3720 \\
        & February 8, 2007 & 54139.82\tablenotemark{a} & & & 3720 \\
        & February 15, 2007 & 54146.33\tablenotemark{a} & & & 3600 \\
        & March 2, 2007 & 54161.51\tablenotemark{a} & & & 2400 \\
        & April 14, 2007 & 54204.59\tablenotemark{a} & & & 3180 \\
        & October 7, 2007 & 54380.10\tablenotemark{a} & & & 6000 \\
        \hline
        AX~J1845$-$0258 & November 14, 2006 & 54053.01 & 18:44:55 & $-$02:56:56 & 3600 \\
        \hline
        GRB~050925 & November 14, 2006 & 54053.09 & 20:13:47 & 34:19:55 & 3600 \\
        & November 17, 2006 & 54056.73 & & & 2760 \\
        \hline
        SGR~1806$-$20 & March 31, 2007 & 54190.47 & 18:08:40 & $-$20:24:37 & 4500 \\
        & April 1, 2007 & 54191.48 & & & 3720 \\
        & April 21, 2007 & 54211.47 & & & 4260 \\
        \hline
        SGR~1900+14 & November 14, 2006 & 54053.05 & 19:07:15 & 09:19:03 & 3600 \\
        \hline
\enddata
\tablenotetext{a}{Observation is part of Target-of-Opportunity observation triggered after an X-ray burst was detected on MJD 54138 using \emph{RXTE} \citep{gdk11}.}
\end{deluxetable}
\end{center}

\begin{center}
\begin{deluxetable}{lccc}
\tablecaption{Rotational Ephemerides Used for Folding. \label{tab:ephems}}
\tablewidth{0pt}
\startdata
        \hline\hline
        Parameter & 1E~1841$-$045 & 1E~2259+586 & 4U~0142+61 \\
        \hline
        Frequency (Hz) & 0.084863091(5) & 0.14328613(6) & 0.1150918267(8) \\
        Frequency derivative (10$^{-15}$ Hz s$^{-1}$) & $-$289(2) & $-$7(5) & $-$26.89(6) \\
        Epoch (MJD) & 54053 & 54050 & 53919 \\
        Reference & \cite{dkg08} & \cite{dib09} & \cite{dkg07} \\
        \hline
\enddata
\tablecomments{Uncertainties reported are the 1 $\sigma$ uncertainties produced by \tempo. Ephemerides are updated versions of what is reported in the references listed, and are valid only for short time intervals surrounding the observation epochs presented in this work.}
\end{deluxetable}
\end{center}

\begin{center}
\begin{deluxetable}{lcccccc}
\tablecaption{Summary of Relevant Properties of the Sources Observed. \label{tab:sourcedetails}}
\startdata
        \hline\hline
        Source & Dist.\tablenotemark{a} & DM & Max DM & SNR Assoc. & $S_{SNR,~1950}$ & $T_{sky}$~\tablenotemark{b}\\
        & & Estimate & Searched & & & \\
        & (kpc) & (cm$^{-3}$ pc) & (cm$^{-3}$ pc) & &(Jy) & (K)\\
        \hline
        \multicolumn{7}{c}{\emph{Confirmed Magnetars}} \\
        \hline
        1E~1841$-$045 & 8.5 & 800 & 2517 & Kes~73 & 3.8 & 8 \\
        1E~2259+586 & 4.0 & 150 & 1007 & CTB~109 & 15.8 & 4 \\
        4U~0142+61 & 3.6 & 100 & 503 & -- & -- & 3.5 \\
        SGR~1806$-$20 & 8.7 & 750 & 2517 & -- & -- & 10 \\
        SGR~1900+14 & 15 & 700 & 2014 & -- & -- & 5.5 \\
        \hline
        \multicolumn{7}{c}{\emph{Magnetar Candidates}} \\
        \hline
        AX~J1845$-$0258 & 8 & 750 & 2517 & G29.6+0.1 & 1.1 & 8.5 \\
        GRB~050925 & 8.8 & 300 & 1008 & -- & -- & 4 \\
        \hline
\enddata
\tablenotetext{a}{The distances reported here are the values used for estimating the DM, as well as for estimating luminosity limits later in this work.}
\tablenotetext{b}{Sky temperatures include a 2.73 K contribution from the cosmic microwave background, as well as a contribution from Galactic synchrotron radiation taken from the \cite{hssw82} all-sky 408 MHz map. Temperatures here are reported for 1950 MHz assuming a power-law spectrum with a synchrotron index of $-$2.5 \citep{kfl+11}.}
\end{deluxetable}
\end{center}

\begin{center}
\begin{deluxetable}{lcccccc}
\tablecaption{Upper Limits on Radio Emission from Magnetars and Magnetar Candidates. \label{tab:uplim}}
\startdata
        \hline\hline
        Source & Epoch & $S_{min}$\tablenotemark{a} & ${L_{min}}$\tablenotemark{a}~\tablenotemark{b} & $S_{min}$ single\tablenotemark{c} & $L_{min}$ single\tablenotemark{b}~\tablenotemark{c} & $S_{min}$ blind\tablenotemark{d}\\
         & & & & & & \\
         & (MJD) & (mJy) & (mJy kpc$^2$) & (mJy) & (Jy kpc$^2$) & (mJy)\\
        \hline
        1E~1841$-$045 
         & 54052.97                  & 0.0102 & 0.74 & 44.1 & 3.2 & 0.0173 \\
        \hline
        1E~2259+586
         & 54053.14                  & 0.0130 & 0.21 & 70.0 & 1.1 & 0.0220 \\
         & 54059.23\tablenotemark{e} & 0.0116 & 0.19 & 69.5 & 1.1 & 0.0198 \\
         & 54059.29\tablenotemark{e} & 0.0108 & 0.27 & 70.1 & 1.1 & 0.0186 \\
        \hline
        4U~0142+61
         & 54053.18                  & 0.0073 & 0.09 & 30.6 & 0.4 & 0.0123 \\
         & 54056.71                  & 0.0115 & 0.15 & 30.4 & 0.4 & 0.0191 \\
         & 54059.38\tablenotemark{e} & 0.0045 & 0.06 & 30.3 & 0.4 & 0.0078 \\
         & 54059.48\tablenotemark{e} & 0.0058 & 0.08 & 31.8 & 0.4 & 0.0099 \\
         & 54138.95                  & 0.0058 & 0.08 & 31.4 & 0.4 & 0.0098 \\
         & 54139.82                  & 0.0063 & 0.08 & 32.9 & 0.4 & 0.0106 \\
         & 54146.33                  & 0.0076 & 0.10 & 30.9 & 0.4 & 0.0128 \\
         & 54161.51                  & 0.0078 & 0.10 & 32.8 & 0.4 & 0.0131 \\
         & 54204.58                  & 0.0069 & 0.09 & 33.7 & 0.4 & 0.0117 \\
         & 54380.10                  & 0.0048 & 0.06 & 32.5 & 0.4 & 0.0082 \\
        \hline
        AX~J1845$-$0258 
         & 54053.01                  & 0.0092 & 0.66 & 38.4 & 2.8 & 0.0125 \\
        \hline
        GRB~050925
         & 54053.09                  & 0.0097 & 0.75 & 30.9 & 2.4 & 0.0097 \\
         & 54056.73                  & 0.0130 & 1.01 & 33.3 & 2.6 & 0.0130 \\
        \hline
        SGR~1806$-$20
         & 54190.47                  & 0.0071 & 0.54 & 37.1 & 2.8 & 0.0121 \\
         & 54191.48                  & 0.0078 & 0.59 & 37.2 & 2.8 & 0.0133 \\
         & 54211.48                  & 0.0069 & 0.52 & 37.0 & 2.8 & 0.0117 \\
        \hline
        SGR~1900+14 
         & 54053.05                  & 0.0071 & 1.60 & 33.8 & 7.6 & 0.0121 \\
        \hline
\enddata
\tablenotetext{a}{A 5\% duty cycle is assumed.}
\tablenotetext{b}{Distances assumed are listed in Table \ref{tab:sourcedetails}.}
\tablenotetext{c}{The values reported here are for a duration of 10 ms. Single pulse durations searched are in the range 0.33 ms to 200 ms.}
\tablenotetext{d}{Blind periodicity search. Assumed 5\% duty cycle, 100 ms period, and DM = 100 cm$^{-3}$ pc.}
\tablenotetext{e}{The observation was split into smaller portions that were searched independently due to significant changes in the data mean caused by RFI (see \S 3).}
\end{deluxetable}
\end{center}

\begin{center}
\begin{deluxetable}{lccc}
\tablecaption{Upper Limits on Single Pulse Rates of Magnetars and Magnetar Candidates. \label{tab:sprate}}
\startdata
    \hline\hline
    Source & Total usable time & Minimum detectable & Single pulse rate limit \\
     & & single pulse luminosity, $L_{1950}$ & \\
     & (hr) & (Jy kpc$^2$) & (hr$^{-1}$) \\
    \hline
    1E~1841$-$045 & 0.66   & 3.2 & 4.5  \\
    1E~2259+586 & 3.5      & 1.1 & 0.85 \\
    4U~0142+61 & 8.5       & 0.4 & 0.35 \\
    AX~J1845$-$0258 & 0.98 & 2.8 & 3.1  \\
    GRB~050925 & 1.6       & 2.4 & 1.9  \\
    SGR~1806$-$20 & 2.8    & 2.8 & 1.1  \\
    SGR~1900+14 & 0.74     & 7.6 & 4.1  \\
    \hline
\enddata
\tablecomments{The single pulse rate limit applies to pulses brighter than the limiting $L_{1950}$ and with durations between 0.33 ms and 200 ms.}
\end{deluxetable}
\end{center}

\clearpage

\end{document}